\documentclass[prl,10pt,twocolumn]{revtex4}
\pdfoutput=1
\usepackage{amsfonts,amsmath}
\usepackage{float}
\usepackage{graphicx}   
\usepackage{ulem}
\usepackage{braket}
\usepackage{verbatim}
\usepackage{mathtools}
\usepackage{times}
\usepackage{setspace}
\usepackage{amsfonts,amsmath,amssymb}
\usepackage{mathtools}
\usepackage{color}
\usepackage{lipsum}

\newcommand{\lac}{\textit{lac}}

\renewcommand{\subsection}[1]{\noindent\textbf{#1. }}

\begin{document}
\title{Switching off: the phenotypic transition to the uninduced state of the lactose uptake pathway}

\author{Prasanna M. Bhogale$^{1}$,
Robin A. Sorg$^{2\#}$,
Jan-Willem Veening$^{3}$,
Johannes Berg$^{1*}$
}

\address{$^1$University of Cologne, Institute for Biological Physics,\\Z\"{u}lpicher Stra{\ss}e 77, 50937 K\"{o}ln, Germany\\
$^2$Molecular Genetics Group, Groningen Biomolecular Sciences and Biotechnology Institute, Center for Synthetic Biology, University of Groningen, Nijenborgh 7, 9747 AG, Groningen, The Netherlands\\
$^3$Department of Fundamental Microbiology, Faculty of Biology and Medicine, CH-1015 Lausanne, Switzerland \\
$^{\#}$ current address IFF Health \& Biosciences, Willem Einthovenstraat 4, 2342 BH Oegstgeest, The Netherlands\\
$^*$ correspondence to \texttt{bergj@uni-koeln.de}
}

\begin{abstract}
The lactose uptake-pathway of \textit{E. coli} is a paradigmatic example of multistability in gene-regulatory circuits. In the induced state of the \lac-pathway,
the genes comprising the \lac-operon are transcribed, leading to the production of proteins which import and metabolize lactose. In the uninduced state, a stable repressor-DNA loop frequently blocks the transcription of the \lac-genes. Transitions from one phenotypic state to the other are driven by fluctuations, which arise from the random timing of the binding of ligands and proteins. This stochasticity affects transcription and translation, and ultimately molecular copy numbers. Our aim is to understand the transition from the induced to the uninduced state of the \lac-operon. We use a detailed computational model to show that repressor-operator binding/unbinding, fluctuations in the total number of repressors, and inducer-repressor binding/unbinding all play a role in this transition. Based on the timescales on which these processes operate, we construct a minimal model of the transition to the uninduced state and compare the results with simulations and experimental observations. The induced state turns out to be very stable, with a transition rate to the uninduced state lower than  $2 \times 10^{-9}$ per minute. In contrast to the transition to the induced state, the transition to the uninduced state is well described in terms of a 2D diffusive system crossing a barrier, with the diffusion rates emerging from a model of repressor unbinding.
\end{abstract}

\maketitle




\section*{Introduction}

Multistable gene regulatory circuits play an important role in diverse biological processes like embryo development \cite{dev1,dev2}, viral reproduction \cite{HIVweinberger,pmid14695251}, and nutrient uptake in bacteria \cite{arabinonsw, xieonmod}. Despite the term ``multistable", regulatory circuits are never truly multi\textit{stable}, as transitions from one phenotypic state to another occur due to random fluctuations: The processes involved in gene expression regulation (changes in operator state, transcription, and translation) are intrinsically stochastic, as is ligand binding. Another source of stochasticity affecting molecular copy numbers is cell division, leading to the random partitioning of molecules in the two daughter cells. These fluctuations can take the system from one (long-lived) phenotypic state to another. The particular fluctuations driving a phenotypic transition vary from system to system. For instance, the transition to competence in \textit{B. subtilis} is driven by fluctuations in the numbers of the ComK protein~\cite{pmid17569828}, while in the arabinose uptake pathway it is entirely driven by the initial distribution of pump proteins in the cell~\cite{arabinonsw}.

\begin{figure*}[bt]
\begin{center}
\includegraphics*[width=.75\textwidth]{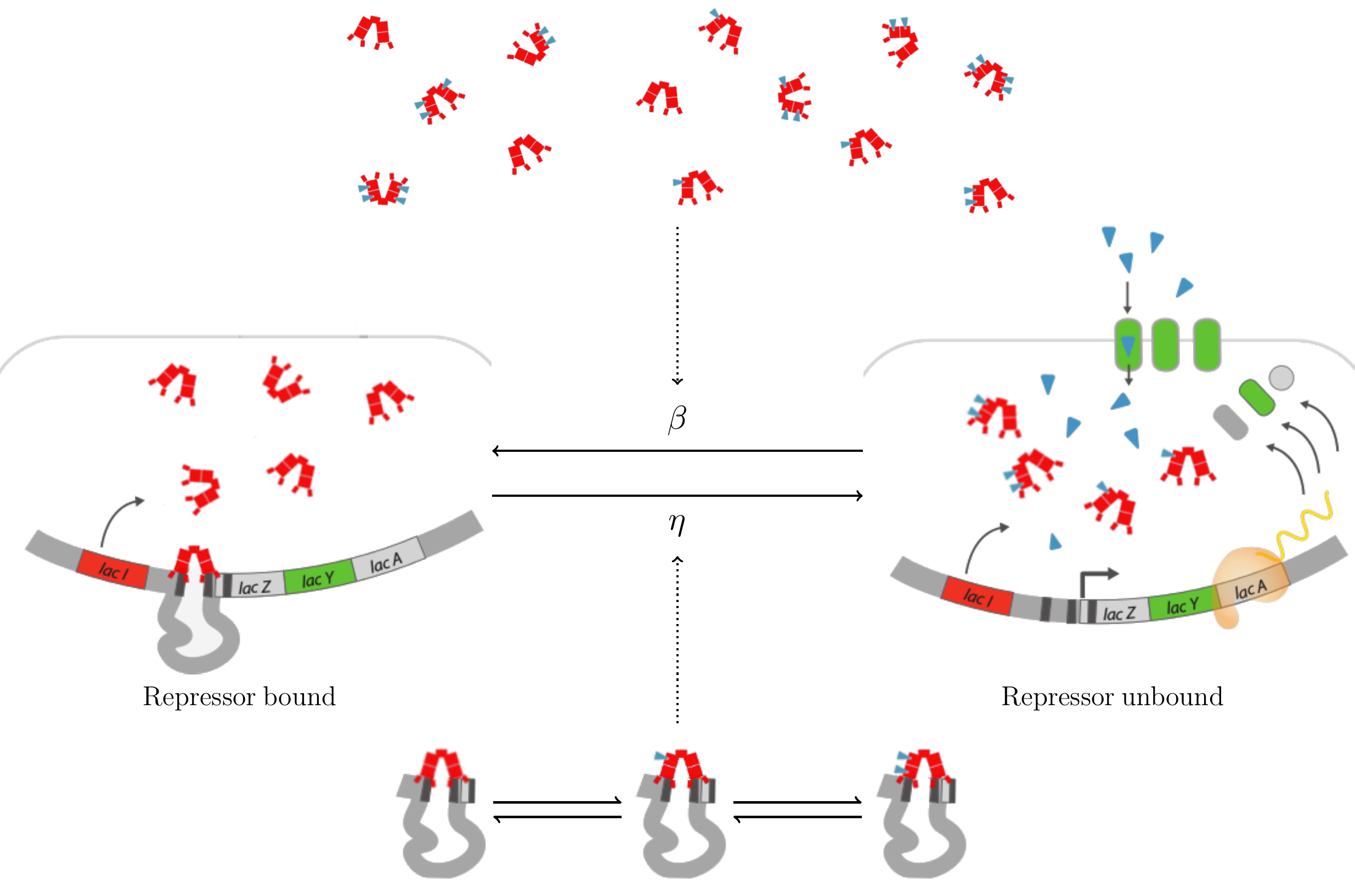}
\caption{{\bf Repressor binding and unbinding.}
The \lac-repressor (red) binds the \lac~regulatory region with two of its `legs' forming a DNA loop that prevents transcription of the \lac~ genes (center left). Without a repressor bound, the~\lac~ genes are expressed, and the pumps LacY (green) import the inducer lactose (blue triangles). The rate of repressor binding $\beta$ depends on the number of repressors present in the cell (top), the rate of unbinding $\eta$ depends on the number of inducers bound to the DNA bound repressor (bottom).}
\label{fig:minsyst}
\end{center}
\end{figure*}

In this paper we focus on the lactose uptake pathway (\lac-pathway), which has been studied extensively since the 1960s~\cite{jacobmonod, novic, multavo, metozvo, pmid17525339, dunwaynum, mullerhillbook, xieonmod, pbhogalelac1,oehlerdistri}. Following early discoveries by Jacob and Monod~\cite{jacobmonod}, and Novick and Weiner~\cite{novic}, the \lac-pathway has become a paradigmatic system for study of gene regulation~\cite{dunwaynum,mullerhillbook}.
The last decade has seen renewed interest in the \lac-pathway due to its multistability, that is the ability of the pathway to sustain different concentrations of a particular protein (or sets of proteins) for long times~\cite{multavo,metozvo,pmid17525339,xieonmod,pbhogalelac1,oehlerdistri}. The induced and uninduced states define two distinct phenotypes differing in their lactose uptake. The induced state persists due to feedback; the more lactose is present in the cell, the more \lac-repressors have a reduced affinity for the \lac~regulatory region and the more pumps are produced and import further lactose. Nevertheless, the number of pumps fluctuates stochastically and will (with a small probability per unit time) cross a threshold to the uninduced state.

In previous work, we have combined a detailed mechanistic model of the \lac-pathway with flow-cytometry experiments to understand the transition from the uninduced to the induced state~\cite{pbhogalelac1}. We found that this transition
occurs when the \lac-repressor unbinds from the \lac~regulatory region and remains unbound for a period of time sufficient to import enough lactose to deactivate all repressors. A simple model of this process gives a transition rate to the induced state which decays exponentially with the inverse of the repressor concentration. This result is borne out experimentally very precisely and over several orders of magnitude (see Fig. 5 of~\cite{pbhogalelac1}).

However, the reverse transition (from the induced to the uninduced state) is still poorly understood.
Our goal here is to identify the sources of stochasticity which affect the transition to the uninduced state and to set up a model of the \lac-pathway which incorporates only the stochastic elements relevant to this transition.
With this approach, we aim to bridge the gap between simple models of gene regulatory networks, which are tractable using the tools of stochastic processes and statistical physics~\cite{keplerandelston,walczakwolynes,XIEPhysRevLett,tsetoggle,aursne1,MorelliTanaseAllenWolde2008, Assaf1, wingreenexp} on the one hand, and numerical simulations of detailed biochemical models on the other hand~\cite{Robertsetal2010,compdetsto,pmid17351004,Santillan2008, Read2015, Ioannis2015
}.

We start with a detailed model of the \lac-pathway to identify the key processes driving the transition to the uninduced state. We find that several sources of stochasticity jointly drive the transition, namely repressor-operator binding/unbinding, inducer-repressor binding/unbinding, and fluctuations in the total number of repressors (ordered in decreasing contribution to the transition).
Based on these processes and the timescales they operate on, we construct a minimal model of the transition from the induced to the uninduced state of the \lac-pathway and compare the results with simulations and experimental observations. We find that the transition to the uninduced state is well described by a model of two-dimensional diffusion across a barrier, and describe how the parameters of the diffusion model emerge from the detailed dynamics of \lac-repressor binding and unbinding. Three techniques turn out to be useful in this context; `smoothing' a stochastic trajectory to identify the key fluctuations in a transition~\cite{pbhogalelac1}, generating Fokker-Planck equations to approximate discrete stochastic processes~\cite{keplerandelston}, and exactly solving the master equation for suitable subsystems of a complex model to derive effective rates.


\section*{Materials and methods}

The \lac~pathway is formed by the \lac~genes \textit{lacY, lacZ}, and \textit{lacA}, which are under joint regulatory control, thus forming a so-called operon. In the induced state of the \lac-pathway, lactose (`inducer') is imported across the cell membrane by the LacY protein (`pumps') and metabolized by the enzyme LacZ into glucose and galactose. Allolactose, a lactose variant originating from LacZ activity, binds to the repressor of the \lac~genes and drastically reduces the affinity of the \lac-repressor to its DNA binding sites \cite{mullerhillbook, xieonmod}. The \lac~ repressor is formed by a dimer of LacI dimers which are expressed constitutively. The reduced affinity causes the repressor to unbind from the \lac~regulatory region, enabling the transcription of the \lac~genes and production of the Lac proteins and further import of lactose.  In the uninduced state on the other hand, LacI frequently binds to two DNA sites in the regulatory region of the \lac-operon and forms a DNA-repressor loop that effectively blocks transcription of the \lac-genes and thus the import of lactose into the cell. For a graphical representation, see Fig.~\ref{fig:minsyst}.

\subsection{A detailed mechanistic model of the \lac-pathway} Our model describes the transcription and translation of mRNA and protein, both of LacY proteins (lactose importer or `pumps') and of LacI proteins (repressors), repressor binding to DNA at its binding sites, DNA looping, the uptake of lactose (inducer) or its analog into a cell, and the passive diffusion of inducers into the cell.
For details, see SI Section 1 and
\cite{pbhogalelac1}.  Almost all rates of these processes are taken from the experimental literature (see SI Table I).
The exceptions are the Michaelis constant of inducer import by LacY, which we use to calibrate our model against the experimental measurements of switching rates from the uninduced to the induced state presented in \cite{pbhogalelac1}, and some of the dissociation rates of DNA and repressors, which we determine using the principle of detailed balance (see SI Section 1.1 and \cite{pbhogalelac1} for details). So although the number of parameters of the model is large, none of the parameters are fitted to data on the transition to the uninduced state.

We performed stochastic simulations of the detailed mechanistic model using the Gillespie algorithm \cite{gillalgo1}. To measure the mean first passage time (MFPT) from the induced to the uninduced state at a given external inducer concentration, previously induced cells (approximately $10^4$ pump proteins at time $t=0$) are simulated with that inducer concentration until the pump protein number crosses a target corresponding to the uninduced state ($\mathcal{O}(100)$).

To pinpoint the relevant fluctuations affecting this transition, we used a smoothing procedure introduced in~\cite{pbhogalelac1}. This procedure reduces the amplitude of fluctuations of a particular component in the pathway, allowing the identification of those fluctuations which influence the rate of transitions to the uninduced state. The smoothing is based on decreasing the step size of a particular reaction, while simultaneously increasing its rate by the same factor (`one tenth of a molecule produced at ten times the rate'). This leaves the mean number of a particular molecule unaffected, but reduces the variance, see~\cite{pbhogalelac1}.

\subsection{{Experimental determination of transition rates to the uninduced state}}
We used the \textit{E. coli} strain  CH458 which has a green fluorescent protein (GFP) gene cassette inserted just after the \lac-genes. Since the GFP gene is co-located with the \lac-genes, the number of GFP and \lac-proteins in the cell are correlated.
High levels of fluorescence indicate cells in the induced state, while uninduced cells have low levels of fluorescence.

We used TMG, a non-metabolizable analog of lactose as an inducer.  We exposed populations of previously induced cells to different concentrations of the inducer and used flow cytometry to determine the fraction of cells that had switched to the uninduced state at a particular later time. Fitted to an exponential decay, measurements at different times yield the transition rate from the induced to the uninduced state at a given concentration of inducer (see SI Section 2 and \cite{pbhogalelac1} for details).

\section*{Results}

In numerical simulations of our mechanistic model, we applied the smoothing procedure to all constituents of the \lac-pathway and their binding states in turn. We found that repressor-operator binding/unbinding, inducer-repressor binding/unbinding, and fluctuations in the total number of repressors all affect the rate of transitions to the uninduced state. Thus, several processes taking place on different timescales are involved in the transition. On the other hand, fluctuations in pump numbers due to translation bursts do not significantly affect the transition rate.

These results are compatible with the following picture of the transition to the uninduced state: The transition to the uninduced state occurs when the copy number of pumps becomes sufficiently low.
This happens as the result of a series of prolonged periods when the \lac-genes are transcriptionally inactive, interspersed with shortened periods when the \lac-genes are transcribed~\cite{keplerandelston, walczakwolynes, MorelliTanaseAllenWolde2008}. Fluctuations to high numbers of repressors lead to fast repressor binding and shorten the transcriptionally active periods. The length of inactive periods, on the other hand, is determined by the rate of repressor unbinding. Repressor unbinding is itself a multi-step process, since the repressor can bind the regulatory region at two
sites simultaneously, and its affinity to these sites changes with the number of inducers bound to the repressor. Fluctuations in the number of inducers bound to a repressor thus affect how long it takes for this repressor to unbind from the regulatory region.

Based on this picture, we propose a minimal quantitative model of the transition to the uninduced state: On timescales which are long compared to the the time intervals between binding/unbinding of repressors, but short compared to the mean first passage time, the number of pumps is described as a diffusive process~\cite{keplerandelston}.  Drift and diffusion of this process are affected by the rate of repressor binding and repressor unbinding.  In the following sections, we discuss this diffusion approximation for the pump copy number, calculate the repressor unbinding rate and how it depends on the inducer concentration, describe the fluctuations in the number of repressors in terms of a second diffusive process. Putting these elements together leads to a Fokker--Planck equation in two variables, which describes the stochastic dynamics of pump and repressor numbers. We calculate the mean-first passage time from the induced state to the uninduced state under this Fokker--Planck equation and compare the results to simulations of the full mechanistic model as well as single-cell experiments.

\subsection{Production of pump proteins} 
\begin{figure}[tb!]
\begin{center}
\includegraphics*[width=.45\textwidth]{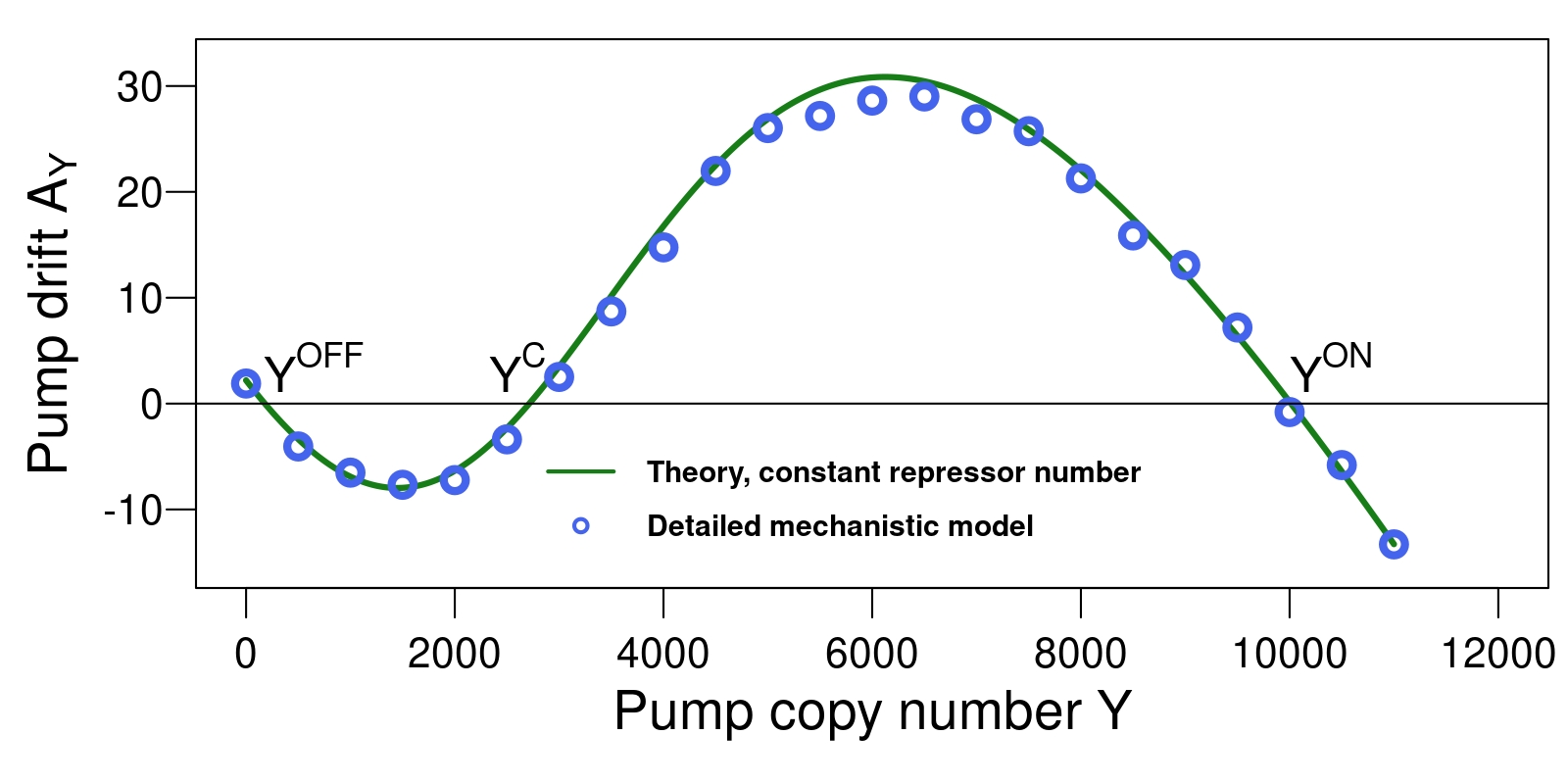}
\includegraphics*[width=.45\textwidth]{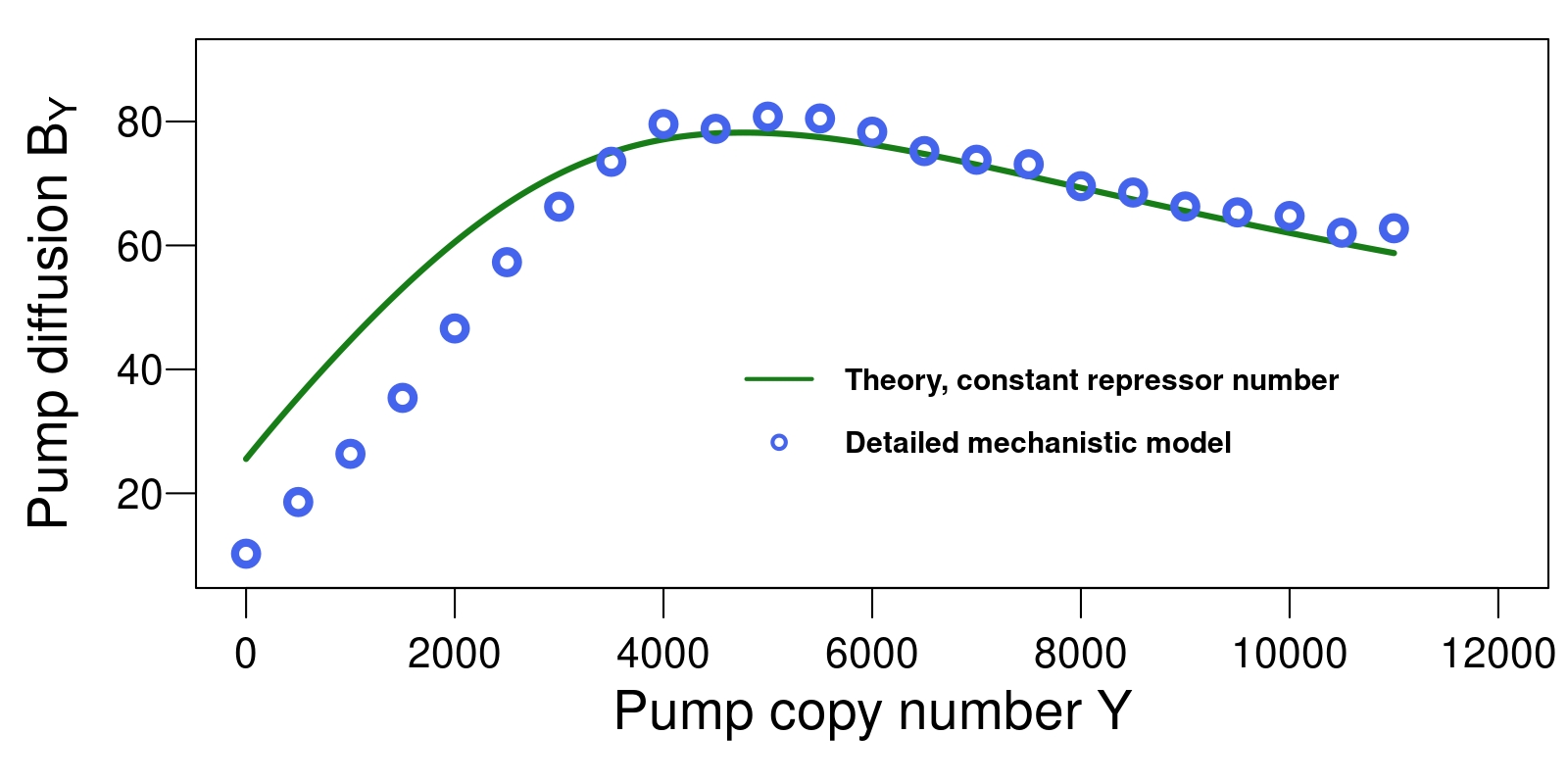}
\caption{{\bf Protein drift and diffusion}. With repressor numbers held constant, protein dynamics can be approximated by a univariate Fokker--Planck equation, Eq. \eqref{eq:profp}. In the top panel, we plot the drift for pump proteins in simulations of our detailed mechanistic model of the \lac~ operon (at an external inducer concentration of 15$\mu$M where the system is still bistable) along with the theoretical drift term from our calculations, Eq. \eqref{eq:prodri} (without any parameter fitting). We observe a very good match between the detailed mechanistic model and the results of \eqref{eq:prodri}. The points of zero drift, $Y^{\text{ON}}$ and $Y^{\text{OFF}}$,  represent the stable points of the induced and uninduced states respectively, while $Y^{\text{C}}$ is the separatrix between them. In the panel below, we plot the diffusion from simulations of our detailed mechanistic model along with the theoretical diffusion term from Eq. \eqref{eq:prodif}. The agreement is good at intermediate and high pump copy numbers, while at low pump copy numbers some discrepancy arises.
}
\label{fig:prodnd}
\end{center}
\end{figure}

We start with a master equation for protein production with repressors binding and unbinding from the operator at rates $\beta$ and $\eta$ respectively~\cite{keplerandelston}.
This is a standard model of gene regulation; later we will add the dependencies of $\beta$ and $\eta$
on repressor numbers and inducer numbers which are specific to the \lac~pathway. $p^\mathbf{0}_{Y}(t)$ denotes the probability that the \lac-genes are transcriptionally inactive (the operator is bound by a repressor, a state denoted $\mathbf{0}$), and there are ${Y}$ pump proteins in the cell, while $p^\mathbf{1}_{Y}(t)$ denotes the probability that the \lac-genes are transcriptionally active (the operator is free of a repressor; state $\mathbf{1}$) and there are ${Y}$ pump proteins in the cell.
The production rates of the pumps when repressor is bound and unbound from operator are denoted by $\bar\xi$ and $\xi$, respectively (see SI Table. I), while the rate at which the pump number is reduced through degradation and cell division is denoted by $\varphi$. The master equation of this process is
\begin{align}
\frac{dp^\mathbf{0}_{Y}}{dt} &= -\left( {Y}\varphi +\bar \xi + \eta \right)p^\mathbf{0}_{Y} + ({{Y+1}})\varphi p^\mathbf{0}_{{{Y+1}}} + \bar \xi p^\mathbf{0}_{{{Y-1}}} + \beta p^\mathbf{1}_{Y} \label{eq:promas1} \\
\frac{dp^\mathbf{1}_{Y}}{dt} &= -\left( {Y}\varphi +\xi + \beta \right) p^\mathbf{1}_{Y} + ({{Y+1}})\varphi p^\mathbf{1}_{{{Y+1}}} + \xi p^\mathbf{1}_{{{Y-1}}} + \eta p^\mathbf{0}_{Y}.  \label{eq:promas2}
\end{align}
To derive a Fokker--Planck equation corresponding to this system of equations, we follow the standard route outlined in \cite{keplerandelston}.
Writing the Taylor series of a function as $f(Y+a) = e^{a\partial_Y}f(Y)$, equations (\ref{eq:promas1}-\ref{eq:promas2}) can be written in terms of the shift operator $e^{\partial_Y}$ and the probability densities $p^{\mathbf{0}}_{Y}$ and $p^{\mathbf{1}}_{Y}$. For fast switching between the operator binding states (compared to the timescales on which the pump numbers change), a quasi steady-state develops between the two operator states. This allows combining the two equations (\ref{eq:promas1}-\ref{eq:promas2}) into a single one for the time evolution of $\rho(Y) = p^{\mathbf{0}}_{Y}+p^{\mathbf{1}}_{Y}$. Restriction to second order terms in the expansion of the shift operator $e^{\partial_Y}$ (the diffusion approximation) gives the Fokker-Planck equation for pump protein dynamics
\begin{align}
{\partial_t\rho(Y)} = -\partial_Y A_Y\rho(Y) +\frac{1}{2}{\partial_Y}^2 B_Y \rho(Y) \label{eq:profp} \ ,
\end{align}
where the drift and diffusion terms are given by
\begin{align}
A_Y &=\varphi(\left<Y\right>-Y) \label{eq:prodri}\\
B_Y &=\frac{\varphi (\left<Y\right>+Y)}{\left<Y\right> }+\frac{2\eta\beta}{(\eta+\beta)^3}(\xi-\bar\xi)^2 \  \label{eq:prodif}
\end{align}
and $\left<Y\right>$ is the mean number of proteins (see SI Table. I). In Fig. \ref{fig:prodnd}, we compare the theoretical drift and diffusion given by equations (\ref{eq:prodri}-\ref{eq:prodif})
with drift and diffusion determined from simulations of our mechanistic model. Crucially, the parameters of this diffusion process depend on the rate of repressor binding $\beta$,  and on the rate of repressor unbinding $\eta$, to which we now turn.

\subsection{Unbinding of the repressor} The unbinding of the repressor LacI from the regulatory region of the \lac-operon is a composite event for two reasons: First, the repressor consists of two `legs', both of which need to unbind from their binding sites. Second, the unbinding rate of each leg depends on the number of inducers bound to the repressor, and that number can change even while the repressor is bound to the \lac-operon.

Each \lac~repressor can bind up to $4$ inducer molecules (one per LacI protein, four of which make a single repressor) and the repressor-operator affinity decreases with each successive inducer bound to the repressor. For instance, a repressor with two inducers bound to it will dissociate from one operator site at the rate of $811/\text{min}$ (see SI Table I), while a repressor with no inducers bound to it has a much lower dissociation rate of $2.4/\text{min}$\cite{dunwaynum}. 

\begin{figure*}[bt]
\begin{center}
\includegraphics*[width=0.9\textwidth]{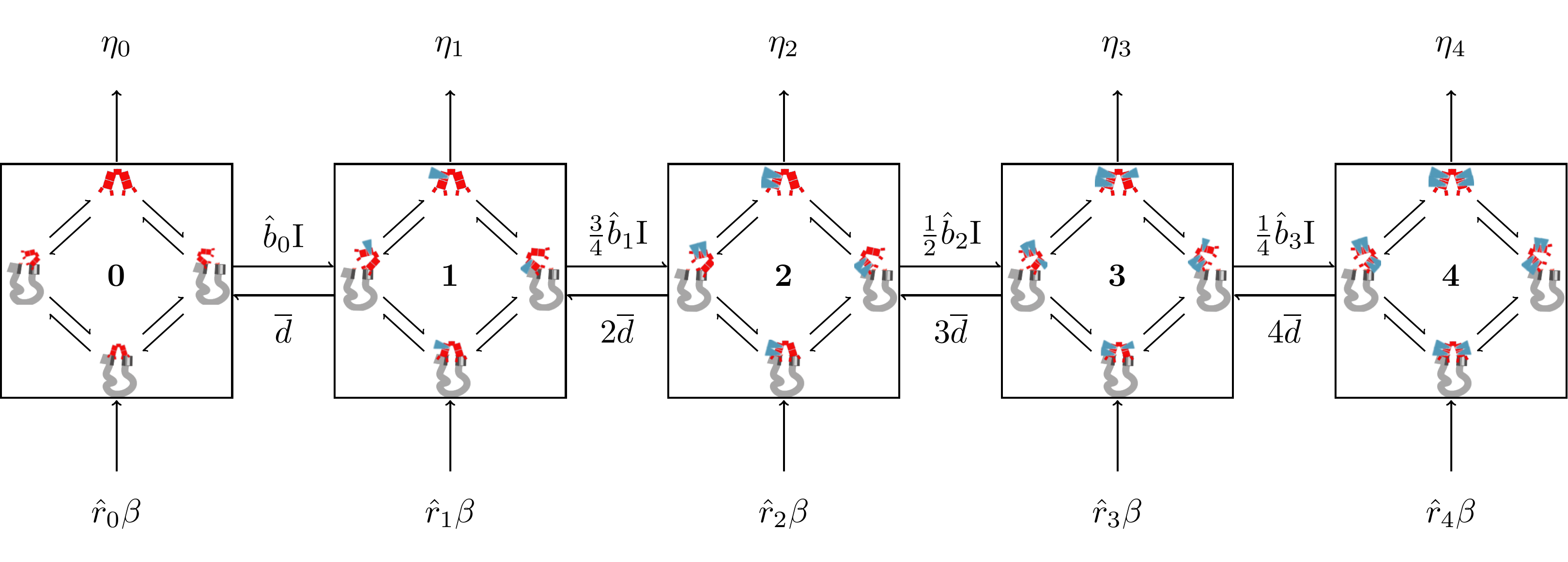}

\caption{{\bf Repressor unbinding.} The repressor unbinding rate depends on the number of inducers bound to the repressor, which changes over time. From left to right, $0,1,2,3$ or $4$ inducers are bound to the repressor, with the changes in the number of inducers are indicated by horizontal arrows. The effective rates of inducer binding to the inducer-DNA system are given by ${\hat{b}_j}$ and the dissociation rate for a single inducer is given by $\overline{d}$ (see SI Table I and SI Section 4.2 for details). The two legs of the repressor can unbind individually (diagonal arrows in each subplot). When both legs have unbound, the repressor dissociates from the regulatory region. The ``exit rate" at which this happens, starting from a given initial number of inducers bound to the repressor, is calculated in SI Section 4.
}

\label{fig:eta}
\end{center}
\end{figure*}

The repressor-operator system can thus be characterized by the number of inducers bound to a repressor, and the number of legs bound to the regulatory region. These configurations are shown in Fig.~\ref{fig:eta}, along with the transitions between them. Unfortunately, not all rates of repressor binding and unbinding in the presence of different numbers of inducers have been measured experimentally (to the best of our knowledge). We use the detailed balance condition to infer some of these rates, see SI Section 1.1 for details.

To compute the effective rate $\eta$ of the unbinding of a repressor-DNA loop (see Fig. \ref{fig:eta}), we first compute the dissociation rate $\{\eta_j\}$ of a repressor with a constant number $j$ of inducers bound to it (SI Section 4). Then we introduce effective rates at which inducers bind to and unbind from that repressor, see Fig. \ref{fig:eta}. The network of stochastic transitions in Fig. \ref{fig:eta} yields a set of linear equations (\ref{eq:tau0}--\ref{eq:tau4}) for the residence times $\{\tau_j\}$ of a repressor on DNA, given that the repressor initially has $j$ inducers bound to it, see SI Section 4 for details. These equations are

\begin{widetext}
\begin{align}
\tau_0 -
\frac{\eta_0}{\left(\eta_0+\hat{b}_0 I\right)^2}
- \frac{\hat{b}_0 I}{\eta_0+\hat{b}_0 I}
\left(\frac{1}{\eta_0+\hat{b}_0 I}+\tau_1\right) &= 0 \label{eq:tau0}\\
\tau_1 -
\frac{\eta_1}{\left(\eta_1+\overline{d}+\frac{3}{4}\hat{b}_1 I\right)^2} -
\frac{\overline{d}}{\eta_1+\overline{d}+\frac{3}{4}\hat{b}_1 I}
\left(\frac{1}{\eta_1+\overline{d}+\frac{3}{4}\hat{b}_1 I}+\tau_0\right) - \nonumber \\
\frac{\frac{3}{4}\hat{b}_1 I}{\eta_1+\overline{d}+\frac{3}{4}\hat{b}_1 I}
\left(\frac{1}{\eta_1+\overline{d}+\frac{3}{4}\hat{b}_1 I}+\tau_2\right) &= 0 \\
\tau_2 -
\frac{\eta_2}{\left(\eta_2+2\overline{d}+\frac{1}{2}\hat{b}_2 I\right)^2} -
\frac{2\overline{d}}{\eta_2+2\overline{d}+\frac{1}{2}\hat{b}_2 I}
\left(\frac{1}{\eta_2+2\overline{d}+\frac{1}{2}\hat{b}_2 I}+\tau_1\right) - \nonumber \\
\frac{\frac{1}{2}\hat{b}_2 I}{\eta_2+2\overline{d}+\frac{1}{2}\hat{b}_2 I}
\left(\frac{1}{\eta_2+2\overline{d}+\frac{1}{2}\hat{b}_2 I}+\tau_3\right) &= 0\\
\tau_3 -
\frac{\eta_3}{\left(\eta_3+3\overline{d}+\frac{1}{4}\hat{b}_3 I\right)^2} -
\frac{3\overline{d}}{\eta_3+3\overline{d}+\frac{1}{4}\hat{b}_3 I}
\left(\frac{1}{\eta_3+3\overline{d}+\frac{1}{4}\hat{b}_3 I}+\tau_2\right) - \nonumber \\
\frac{\frac{1}{4}\hat{b}_3 I}{\eta_3+3\overline{d}+\frac{1}{4}\hat{b}_3 I}
\left(\frac{1}{\eta_3+3\overline{d}+\frac{1}{4}\hat{b}_3 I}+\tau_4\right) &= 0 \\
\tau_4 -
\frac{\eta_4}{\left(\eta_4+4\overline{d}\right)^2} -
\frac{4\overline{d}}{\eta_4+4\overline{d}}\left(\frac{1}{\eta_4+4\overline{d}}+\tau_3\right) &= 0, \label{eq:tau4}
\end{align}
\end{widetext}
where ${\hat{b}_j}$ are the effective rates of inducer binding to the inducer-DNA system and $\overline{d}$ is the dissociation rate for a single inducer.

We compute the effective residence time of a repressor on DNA $\frac{1}{\eta}$ by averaging the $\{\tau_j\}$ obtained by solving equations (\ref{eq:tau0}--\ref{eq:tau4}), over the probabilities that a repressor-DNA loop has $j$ inducers bound to it. These probabilities $\{\hat{r}_j\}$ of a DNA-repressor loop being formed with $j$ bound inducers can be calculated given the relative number of repressors and inducers in the cell and the fact that inducer binding and unbinding is fast compared to the timescale over which the number of repressors change (see SI Section 4.4). The  effective unbinding rate $\eta$ is then given by
\begin{equation}
\eta = \left(\sum_{j=0}^4 \hat{r}_j \tau_j \right)^{-1}. \label{eq:eta}
\end{equation}

In Eqn. (\ref{eq:eta}), the residence times  $\{\tau_j\}$ and the probabilities $\{\hat{r}_j\}$ are functions of the number of inducers in the cell, which in turn is a function of number of pump proteins and external inducer concentration. At a fixed number of pumps, an equilibrium between inducer import and inducer dilution through cell growth is established.  The mean number of inducers $I$ can then be written as a function of pump number $Y$ (see SI Table. I)

\begin{equation}
   {I}({Y})=\frac{m}{\varphi} \frac{E}{E_h + E} Y      \ ,
\label{eq:inducer_protein}
\end{equation}
where $\varphi$ is the dilution rate, $m$ is the rate of inducer import per pump, $E_h$ is the Michaelis constant which we use as a fitting parameter, and $E$ is external TMG concentration. To the best of our knowledge, measurements of the quantities $m$ and $E_h$ are not available for TMG, the lactose analog used in our experiments \cite{pbhogalelac1}. From equation \ref{eq:inducer_protein} it is clear that assumptions about the value of $m$ will strongly affect the fitting value for $E_h$. In our mechanistic model, we use the value of $m$ reported in \cite{importrateexperiment} by Smirnova {et. al} for the sugar NPG (1260/min). We determine the Michaelis constant $E_h$ by fitting simulations of the detailed mechanistic model to experimental data on the switching from the uninduced to the induced state from \cite{pbhogalelac1} (the reverse transition to the one considered here, see SI Section 3 for details). The best fit was obtained for $E_h=1.05\times 10^{5} \mu M$. For comparison, values of the Michaelis constants $E_h$ for lactulose transport by LacY ($2.4\times 10^2\mu M$), and sucrose ($6.7\times 10^3 \mu M$), fructose ($3.5\times 10^4 \mu M$) transport by CscB are reported by Sugihara {et. al} in \cite{sugimp}.
Since there are significant variations between different sugars, the fitted value of $E_h$ for TMG is not implausible, however it is sensitive to other parameters of the model for which TMG-specific measurements are not available. Specifically, for large values of $E_h$, equation \ref{eq:inducer_protein} depends on $m$ and $E_h$ only through their ratio. The value obtained for $E_h$ might thus reflect simply a value of the parameter $m$ that is not correct for the inducer TMG used here. However, while $E_h\gg 100 \mu M$ the ratio of $m$ and $E_h$ will be independent of inaccuracies in the value of $m$.

\begin{figure}[bt]
\begin{center}
\includegraphics*[width=.45\textwidth]{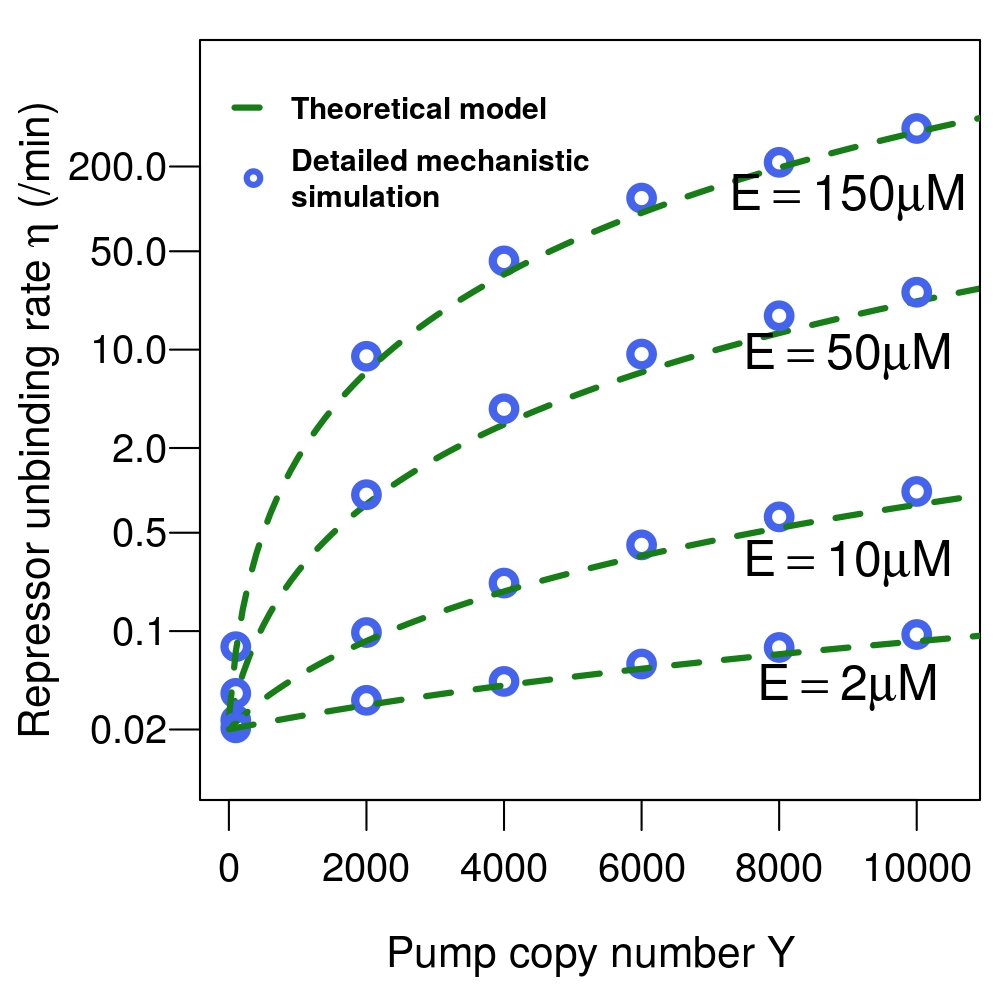}
\caption{{\bf
Rate of repressor dissociation from the operator.}
This figure shows the rate of repressor dissociation from the regulatory region at different copy numbers of the pumps. Pump numbers affect the dissociation rate via the number of inducers, as an inducer-bound repressor dissociates quickly from the regulatory region. Blue circles show the dissociation rate under the detailed mechanistic model, the green line shows the results of equations~(\ref{eq:eta}) and~(\ref{eq:inducer_protein}) without any parameter fitting.}
\label{fig:etacomp}
\end{center}
\end{figure}

Equations~\eqref{eq:eta} and \eqref{eq:inducer_protein} jointly establish a relationship between the number of pumps and the repressor dissociation rate. This is an effective rate, which results from inducers repeatedly binding/unbinding from repressors and influencing the residence time of repressors on the regulatory region. Equations~(\ref{eq:eta}) and (\ref{eq:inducer_protein}) quantify the amount of feedback in the \lac-pathway: The more pumps there are, the faster the \lac-repressor will unbind from the regulatory region, enabling the production of further pumps.  In Fig. \ref{fig:etacomp} we compare the results from Eqns.~\eqref{eq:eta} and~\eqref{eq:inducer_protein} for the effective dissociation rate to the rate at which repressors unbind from the regulatory region in simulations of the detailed mechanistic model and find very good agreement.
This effective repressor dissociation rate will be used below as the rate at which the \lac~operon turns from the transcriptionally inactive to the active state, which enters the diffusion model of the pump numbers.

\subsection{Repressor binding} The second quantity entering the diffusion model of the pump numbers is the rate $\beta$ at which
the regulatory region is bound by a repressor. This rate is proportional to the number of repressors in the cell, the rate $g$ at which repressors can find their binding site on DNA (see SI Table I for details) and the rate at which these repressors can form a repressor-DNA loop, which in turn depends on the numbers of inducers bound to the repressor. The binding rate $\beta$ is a function of the inducer numbers inside the cell via the fractions of free repressors in the cell with $j$ inducers bound to them $\{r_j\}$
\begin{equation}
\beta = gR\sum_{j=0}^4 r_j \frac{c}{c+w_{4-j}}, \label{eq:beta}
\end{equation}
where $R$ denotes the number of repressors in the cell, $c$ is the rate at which a repressor with one leg bound to DNA forms the repressor-DNA loop. $w_{4-j}$ denotes the rate at which a repressor with $j$ bound inducers detaches one leg from its binding site on DNA (see SI Table I). The quantity $R\sum_{j=0}^4 r_j \frac{c}{c+w_{4-j}}$ can be interpreted as the effective number of repressors in the cell.

\subsection{Repressor number fluctuations} 
\begin{figure}[bt]
\begin{center}
\includegraphics*[width=.45\textwidth]{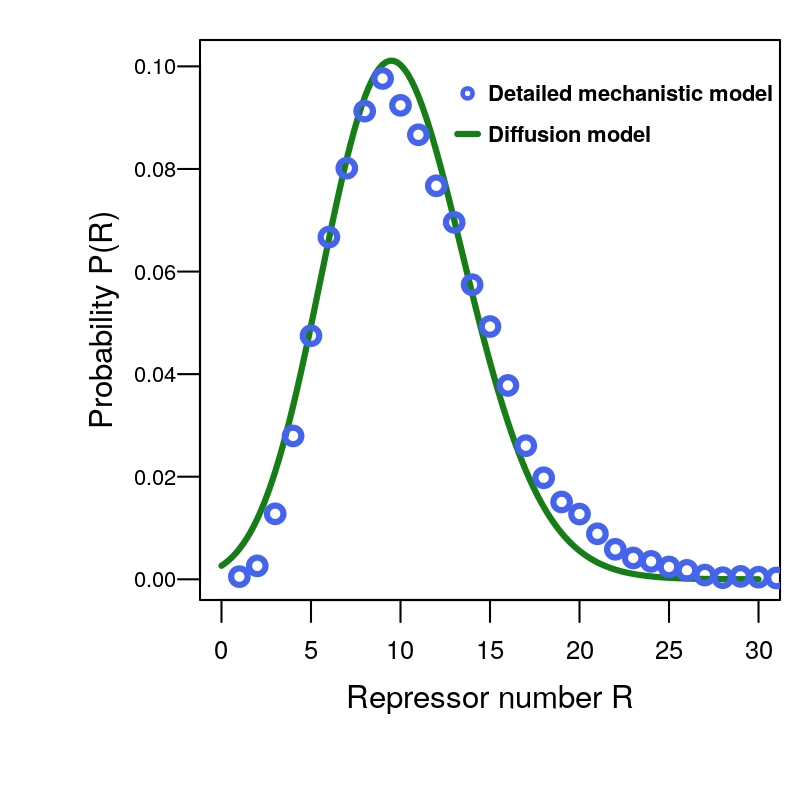}
\caption{{\bf Distribution of repressor numbers.} Repressor number fluctuate since the \textit{lacI}-gene is expressed at such low levels that typically there are none or one copy of \textit{lacI}-mRNA in the cell. Green lines show the probability distribution function from the diffusion model defined by drift~\eqref{eq:repdri} and diffusion~\eqref{eq:repdif}, blue circles give the corresponding quantities observed in the detailed mechanistic model. }
\label{fig:repfluc}
\end{center}
\end{figure}

While the \textit{lacI} gene is expressed constitutively, fluctuations in the number of \textit{lacI}-mRNA lead to fluctuations in the number of repressors. Since the transcription rate $c^{R}$ of the \textit{lacI} gene is much smaller than the mRNA degradation rate $\phi$ (see SI Table. I), we assume that there is at most one $\text{\textit{mRNA}}_{{I}}$ molecule in the cell. This approximation enables setting up a  Fokker--Planck equation for the repressor number using a procedure identical to that used earlier to derive Eq. (\ref{eq:profp}). The resulting drift and diffusion of the repressor numbers $R$ are

\begin{align}
A_R &= s\frac{c^Rl^R}{\phi} - R\varphi , \label{eq:repdri}\\
B_R &= s \frac{c^R\left(l^R\right)^2}{\phi^2} + R\varphi +s^2\frac{2 c^R \phi {\left(l^R\right)}^2}{\left(c^R + \phi\right)^3} , \label{eq:repdif}\
\end{align}
where $\phi$ is the total rate of mRNA dilution and degradation,
while $c^{R}$ and $l^{R}$ are  the transcription and translation rates for the LacI protein (see SI Table. I). The factor $s=1/4$ reflects the fact that repressors consist of $4$ LacI proteins each. Figure \ref{fig:repfluc} compares the distribution of repressor numbers resulting from (\ref{eq:repdri}-\ref{eq:repdif}) to the distribution observed in simulations of our detailed mechanistic model and finds good agreement between them.

\subsection{Setting up the diffusion model}
Drift and diffusion of the pump numbers given by equations~\eqref{eq:prodri} and~\eqref{eq:prodif} depend on the repressor unbinding rate $\eta$ and the binding rate $\beta$.
The unbinding rate depends on the pump copy number via equations~\eqref{eq:eta} and~\eqref{eq:inducer_protein}, and the binding rate fluctuates along with the number of repressors given by equations \eqref{eq:repdri} and \eqref{eq:repdif}. Putting these results together gives a
bivariate Fokker--Planck equation that describes the joint time evolution of protein numbers $Y$ and repressor numbers $R$,
\begin{widetext}
\begin{equation}
{\partial_t\rho(Y,R)} = -\partial_{Y} A_{Y}\rho(Y,R) +\frac{1}{2}{\partial_{Y}}^2 B_{Y} \rho(Y,R) -\partial_{R} A_{R}\rho(Y,R) +\frac{1}{2}{\partial_{R}}^2 B_{R} \rho(Y,R) \label{eq:2Dfp}.
\end{equation}
\end{widetext}

\subsection{Mean first passage times from the induced to the uninduced state}

Starting in the induced state, with pump number $Y$ set equal to $Y^{\text{ON}}=10^4$, the transition to the uninduced state occurs when the number of pump proteins
reaches the vicinity of the zero-drift point defining the uninduced state $Y^{\text{OFF}}$,  with $\mathcal{O}(100)$ pump proteins.
In the absence of repressor fluctuations, the dynamics of the number of pump protein can be described by the univariate Fokker--Planck equation Eq. \eqref{eq:profp}. The values of the protein numbers where the drift Eq. \eqref{eq:prodri} equals zero, ${Y}^{\text{ON}}$ and ${Y}^{\text{OFF}}$, are the stable fixed points of the Fokker--Planck equation in the noiseless (deterministic) limit. In the presence of noise,
they correspond to the long-lived induced and uninduced states respectively, while ${Y}^{\text{C}}$ corresponds to the separatrix between them (the unstable fixed point, see Figure \ref{fig:prodnd}). Such a visualization of the stationary points in terms of zeros of a closed-form drift function is not possible for dynamics of pump proteins in the presence of repressor fluctuations. In this case, the dynamics of the system is described by the two-dimensional Fokker--Plank equation \eqref{eq:2Dfp}.
The mean first passage (MFP) time $\Gamma$ from one stable fixed point $\left(Y^{\text{ON}}\right)$ to another $\left(Y^{\text{OFF}}\right)$ can still be calculated based on the backward Fokker--Planck equation~\cite{keplerandelston,gardiner1985handbook}
\begin{equation}
-1 = A_{Y}\partial_{Y} \Gamma +\frac{1}{2}B_{Y}{\partial_{Y}}^2 \Gamma +A_{R}\partial_{R} \Gamma +\frac{1}{2}B_{R}{\partial_{R}}^2  \Gamma \label{eq:mfpt} \ ,
\end{equation}
subject to the boundary conditions
\begin{align}
\Gamma(Y^{\text{OFF}}) &= 0,\\
\frac{d\Gamma}{dt}\Bigr |_{Y^{\text{ON}}} &= 0 \ .\label{eq:rigidgamma}
\end{align}
These boundary conditions state that mean first passage time is $0$ when the dynamics starts already at the destination $Y^{\text{OFF}}$, while near the induced state $Y^{\text{ON}}$, the first passage time is insensitive to small perturbations in the starting point.  
Inducer numbers
 in the cell at start and end points do not affect the first passage time since they quickly reach the steady state determined by the number of pump proteins
 given by Eq. \eqref{eq:inducer_protein}.

\begin{figure}[bt]
\begin{center}
\includegraphics*[width=.45\textwidth]{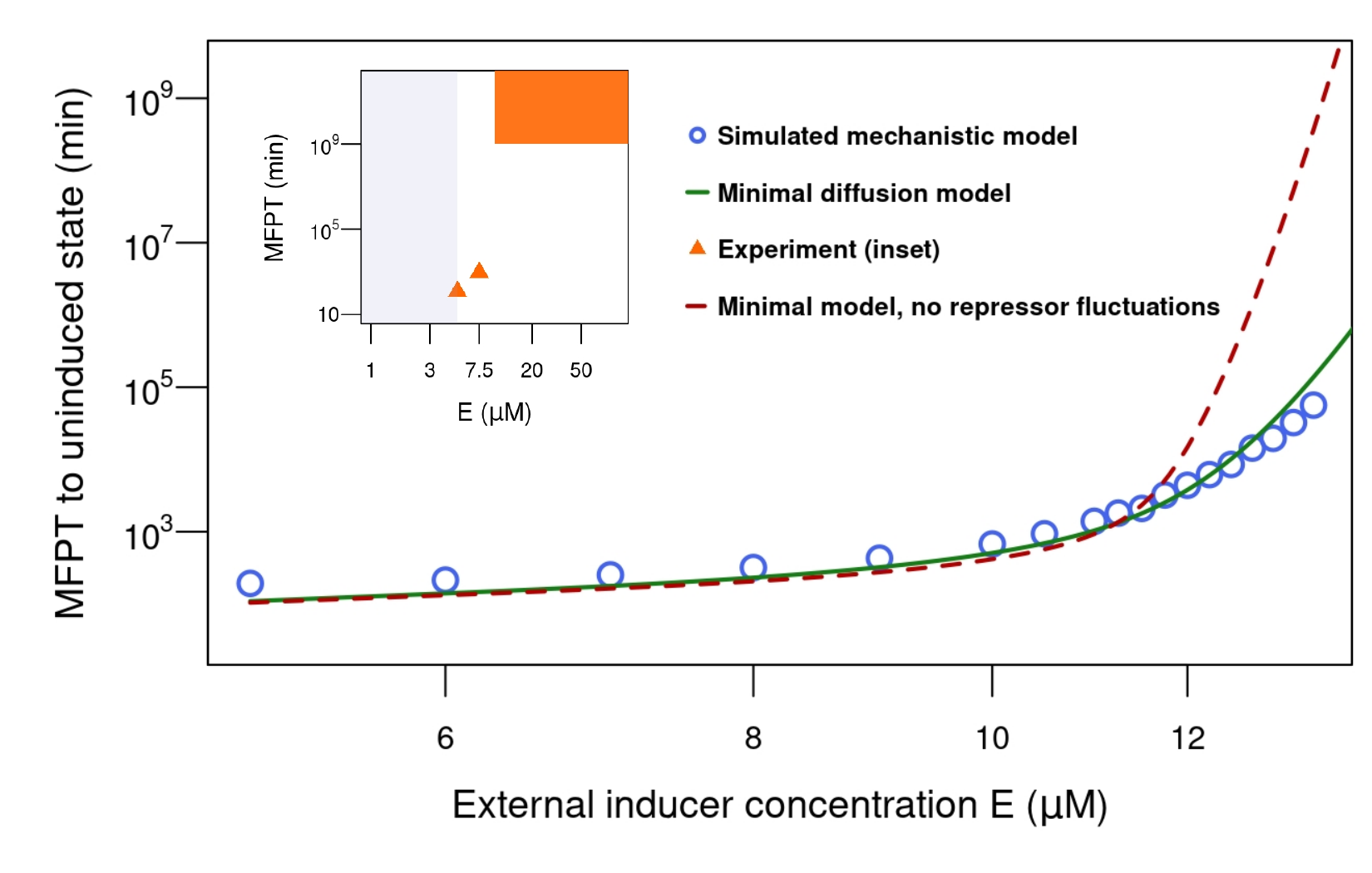}
\caption{{\bf Mean first passage times.} The mean first passage time starting from the induced state with $Y^{\text{ON}}$ pumps to the uninduced state with $Y^{\text{OFF}}$ pumps is plotted against the external inducer concentration.
The results from the minimal diffusion model \eqref{eq:mfpt} agree very well with simulations of the detailed mechanistic model (blue circles and green line, respectively). We also show the results from the detailed mechanistic model when repressor numbers are kept constant (red dashed line). Neglecting repressor fluctuations slows down the transition at high inducer concentrations. The super-exponential growth of mean first passage times with external TMG concentration is consistent with observations from our experiments (inset). The transition can only be observed at a concentrations of $5\mu m$ and $7.5\mu M$  (estimated MFTP of 117 minutes and 920 minutes respectively, see SI Section 2)
due to the
the first passage times for concentrations above $10\mu M$ being too high to be observable in our experiments. The duration of our experiments gives a lower bound of $\mathcal{O}(10^{9})$ minutes for the mean first passage time (see main text). This lower bound is indicated by the rectangle in the top right of the inset.
}
\label{fig:results}
\end{center}
\end{figure}

To determine the mean first passage time, we solve the backward Fokker--Planck equation (\ref{eq:mfpt}) numerically. Fig. \ref{fig:results} shows the results as a function of the external inducer concentration. We compare the first passage times to the uninduced state obtained from Eq.(\ref{eq:mfpt}) to simulations of the detailed mechanistic model and find a good match between the diffusion model and the detailed mechanistic model.
We find that the mean first passage time increases exponentially with external inducer concentration once the induced state is viable. As a result, for concentrations larger than approximately $10\mu M$, the induced state is extraordinarily stable, with mean-first passage times exceeding $10^9$ minutes. The induced state can thus persist over many generations, and is actually transmitted more stably to subsequent generations than genetic information: A generation time of $60$ min implies a transition rate to the uninduced state of $\mathcal{O}(10^{-7})$ per generation, compared to a point mutation rate of $\mathcal{O}(10^{-6})$ per generation. A similar situation has been found in the dormant state of the $\lambda$-phage~\cite{pmid14695251}.

On the other hand, at external inducer concentrations below $5 \mu$M, there is no long-lived induced state, as the lactose that can be imported at such low external concentrations is not sufficient to deactivate all repressors and sustain an induced state. Thus, even if initial pump numbers are large, they quickly decay and the cells collectively transition to the uninduced state. This dynamics has been called a `ballistic transition'~\cite{metozvo}.

Fluctuations in the number of repressors contribute in different ways to the transition to the uninduced state.
Performing simulations of the detailed mechanistic model at constant number of repressors (with repressor number equal to their mean $R=10$ under the full model),
increases the mean first passage time significantly at high inducer concentrations.
Higher-than-average repressor numbers lead to long periods where the \lac-genes are transcriptionally silenced, making it easier for the pump number to reach lower levels, which can effectively lower the barrier to be crossed by diffusion. 

\subsection{Comparison with experiments} Due to the long mean-first-passage time, the transition to the uninduced state is challenging to observe experimentally. Specifically, the rapid increase of the MFPT with external inducer concentration means that the transition can only be observed in a narrow window where the induced state is stable but the MFPT is
shorter than time scale over which the experiment is performed. The experiments were performed as described in~\cite{pbhogalelac1}.
We found the induced state to be unstable at TMG concentrations below $5\mu$M, and observed a `ballistic collapse' of the entire cell population to the uninduced state. At concentrations greater than $10\mu$M, we did not observe any transitions over $8$ hours in a population of approximately $10^6$ cells, implying that the mean first passage time is greater than $2 \times 10^9$ minutes (see Fig \ref{fig:results} inset).  On the other hand, at an intermediate concentration of $7.5\mu$M, we did observe transitions to the uninduced state occurring at a rate of $1.08\times 10^{-3}/$min. This behavior qualitatively matches the numerical simulations. For a quantitative comparison far higher population sizes of cells would be needed to observe more transitions to the uninduced state even when the MFPT is large.

\section{Discussion}

In this paper we have identified the fluctuations that drive the transition to the uninduced state of the \lac~pathway of \textit{E. coli}. The repressor-operator binding/unbinding, inducer-repressor binding/unbinding, and fluctuations in the total number of repressors all contribute to the stochastic transition between these two states. To make this system tractable, we compute effective rates of repressor binding and dissociation from DNA as functions of pump numbers, and use these in a bivariate Fokker-Planck equation that captures the stochastic dynamics of pump and repressor copy numbers. From this equation, we compute the mean first passage time to the uninduced state and compare the result to numerical simulations of a detailed mechanistic model.

The transition mechanism is thus the diffusive crossing of a barrier. This is different from the mechanism we previously found for the reverse transition in the same system, the transition from the uninduced to the induced state. There, the key step turned out to be the unbinding of the \lac-repressor for a time period exceeding a particular critical duration~\cite{pbhogalelac1}. These  different types of mechanisms have previously been discussed~\cite{walczakwolynes} as two distinct possibilities in which gene expression fluctuations can lead to phenotypic switching. It is interesting to see both of them realized in a single, well-studied system.

We find that the barrier-crossing mechanism can give the induced state a remarkable stability: the mean-first-passage-time increases exponentially with the external inducer concentration. As a result, stochastic transitions to the uninduced state can occur at rates as low as $10^{-8}$ per minute, and are thus hard to observe experimentally.
Also, the transition to the uninduced state is much slower than the reverse transition to the induced state~\cite{pbhogalelac1}. A similar asymmetry in the transition rates is found as well in the arabinose uptake network of \textit{E. coli}~\cite{ara2}. However, using the natural inducer lactose in this system would make the induced state less stable (because the inducer is degraded by LacZ), possibly even to the point where the bistability is lost~\cite{zander2017bistability}.

If the imbalance between the rates of the two transitions persists, it clearly limits the usefulness of stochastic transitions between phenotypic states as a `bet hedging' strategy: It has been proposed that bacteria use stochastic transitions to ensure that at any point in time, part of a population of cells is in one state, part in the other. Depending on external conditions, one state will confer a fitness advantage compared to the other; with stochastic transitions between states, part of the population will always be in the advantageous phenotypic state, no matter how external condition change over time (for a review, see~\cite{fraser2009chance}
).
However, if one rate dominates over the other, nearly all cells will be in only one of the two states. Hence, in the case of the \lac~pathway, stochastic transitions between long-lived states may simply be an unavoidable consequence of implementing bi-stability in a system containing stochastic components, not a feature that confers an evolutionary advantage.

\section*{Competing interests}

The authors declare that they have no competing interests.



\section*{Acknowledgements}

We thank  A. Gordon for the \textit{E. coli} strain CH458 and Martina Markus for figure design.

\section*{Funding}

This work was supported by SysMO2 Noisy Strep.  Work in the Veening lab is also supported by the Swiss National Science Foundation (SNSF) (project grant 31003A\_172861), a JPIAMR grant (40AR40\_185533) from SNSF and ERC consolidator grant 771534-PneumoCaTChER, work in the Berg group is supported by DFG grant SFB 1310.

\bibliography{lacbib_v2}
\bibliographystyle{unsrt}

\end{document}